# Roadmap towards the redefinition of the second

N. Dimarcq[1], M. Gertsvolf[2], G. Mileti[3], S. Bize[4], C.W. Oates[5], E. Peik[6], D. Calonico[7], T. Ido[8],

P. Tavella[9], F. Meynadier[9], G. Petit[9], G. Panfilo[9],

J. Bartholomew[10,], P. Defraigne[11], E. A. Donley[5], P. O. Hedekvist[12], I. Sesia[7], M. Wouters[13],

P. Dubé[2], F. Fang[14], F. Levi[7], J. Lodewyck[4], H. S. Margolis[15], D. Newell[5], S. Slyusarev[16], S. Weyers[6], J.-P. Uzan[17], M. Yasuda[18], D.-H. Yu[19],

C. Rieck[12], H. Schnatz[6], Y. Hanado[8], M. Fujieda[8,21], P.-E. Pottie[4], J. Hanssen[20], A. Malimon[16], N. Ashby[5].

1	ARTEMIS, Université Côte d'Azur, Observatoire de la Côte d'Azur, CNRS, Nice, France
2	National Research Council, Ottawa, Canada
3	Université de Neuchâtel, Neuchâtel, Switzerland
4	LNE–SYRTE, Observatoire de Paris, Université PSL, CNRS, Sorbonne Université, Paris, France
5	National Institute of Standards and Technology, Boulder, Colorado, USA
6	Physikalisch-Technische Bundesanstalt (PTB), Bundesallee   Braunschweig, Germany
7	Istituto Nazionale di Ricerca Metrologica, Turin, Italy
8	National Institute of Information and Communications Technology, Koganei, Tokyo, Japan
9	Bureau International des Poids et Mesures, Sèvres, France
10	Emirates Metrology Institute (QCC EMI), Abu Dhabi, United Arab Emirates
11	Royal Observatory of Belgium, Brussels, Belgium
12	Research Institutes of Sweden, Borås, Sweden
13	National Measurement Institute, Sydney, Australia
14	National Institute of Metrology, Beijing, China
15	National Physical Laboratory, Teddington, UK
16	FSUE VNIIFTRI, Mendeleevo, Solnechnogorsky District, Moscow Region, Russian Federation
17	Sorbonne Université, CNRS, UMR 7095, Institut d'Astrophysique de Paris, France
18	National Metrology Institute of Japan, National Institute of Advanced Industrial Science and Technology, Tsukuba, Ibaraki, Japan
19	Korea Research Institute of Standards and Science, Daejeon, Republic of Korea
20	United States Naval Observatory (USNO), Washington, DC, USA
21	National Astronomical Observatory of Japan, 2-21-1 Osawa, Mitaka, Tokyo, Japan

**Abstract**
This paper outlines the roadmap towards the redefinition of the second, which was recently updated by the CCTF Task Force created by the CCTF in 2020. The main achievements and the open challenges related to the status of the optical frequency standards, their contribution to time scales and UTC, the possibility of their comparison and the knowledge of the Earth's gravitational potential at the necessary level of uncertainty are discussed. In addition, the mandatory criteria to be achieved before redefinition and their current fulfilment level, together with the redefinition options based on a single or on a set of transitions are described.

## 1.  Introduction

The definitions of the base units of the International System of Units (SI) [1] are decided by the General Conference on Weights and Measures (CGPM) that supervises the work of the International Committee for Weights and Measures (CIPM) and its Consultative Committees.  Following definitions based on astronomical phenomena, the definition of the SI unit of time, the second, has relied since 1967 on the caesium atom hyperfine transition frequency (Section 2). Caesium primary frequency standards are currently realizing this unit with a relative frequency uncertainty at the low $10^{-16}$ level, but in the last two decades they have been surpassed by optical frequency standards (OFS) showing much lower uncertainties, currently 2 orders of magnitude better.

In 2016, the Consultative Committee for Time and Frequency (CCTF) set up a first version of the roadmap towards the redefinition of the second and the associated conditions for the redefinition [2, 3].

Since June 2020, the roadmap has been updated by a dedicated CCTF Task Force on this topic, with three subgroups related to:

A. Requests from user communities, National Metrology Institutes and Liaisons
B. Atomic frequency standards, and possible redefinition approaches
C. Time and Frequency dissemination and time scales.

The CCTF has gathered feedback on the redefinition of the second through a global consultation of concerned communities and stakeholders, which was carried out through an online survey from December 2020 to January 2021. It has analysed the needs and possible impacts of a new definition, not just scientific and technological, but also regulatory and legislative (Section 3). The choice of the new definition is central to the debate: the CCTF has analysed the various options that can be envisaged and identified the pros and cons of each possibility (Section 4). The CCTF has updated criteria and conditions that quantify the status of the developments and their maturity for a redefinition (Section 5). The fulfilment of mandatory criteria relies on the progress of ultra-low uncertainty and reliable Optical Frequency Standards (OFS - Section 6) and Time and Frequency (TF) transfer and comparison techniques (Section 7) required for the realization of the new definition and its dissemination towards users, including the contribution of OFS to the International Atomic Time scale (TAI).

2. **History of definitions**

Until 1967, the SI definition of time had been based on astronomy. It was initially the fraction 1/86 400 of the mean solar day but observation of unpredictable variations in the Earth rotation rate led in 1960 to a change of the definition to choose a more stable astronomical phenomenon: the motion of the Earth around the Sun, with an SI second equal to the fraction 1/31 556 925.9747 of the tropical year 1900.

Thanks to the rapid progress of caesium thermal beam frequency standards, the SI definition of the second left the field of astronomy in 1967 to enter the field of quantum physics, with the definition exploiting the benefits of high precision frequency measurements [4]. The second became at that time the "the duration of 9 192 631 770 periods of the radiation corresponding to the transition between the two hyperfine levels of the ground state of the caesium 133 atom". In 1999, to take black body radiation shifts into account, an addendum to the initial definition was issued to specify that the definition refers to a caesium atom at rest at a temperature of 0 K.

The 26th meeting of the CGPM (2018) marked an important step with the revision of the SI system of units and the redefinition of four base units, by fixing the values of fundamental constants: kilogram (Planck constant $h$), ampere (elementary charge $e$), kelvin (Boltzmann constant $k_B$), and mole (Avogadro constant $N_A$). The basis of the definition of the SI second remained the same but the wording changed in order to be consistent with the general spirit of the new SI, fixing the value of the caesium frequency: "The second, symbol s, is the SI unit of time. It is defined by taking the fixed numerical value of the caesium frequency $\Delta\nu_{Cs}$, the unperturbed ground-state hyperfine transition frequency of the caesium-133 atom, to be 9 192 631 770 when expressed in the unit Hz, which is equal to s$^{-1}$". In this revised SI, the unit of time has a central position since fixing the values of fundamental constants leads to a direct dependence of all the units, except the mole, on the definition of the second (Table 1).

| Unit | Defining constant | Dependence on other units | s | m | A | kg | K | Cd |
|---|---|---|---|---|---|---|---|---|
| s | $\Delta\nu_{Cs}$: unperturbed ground-state hyperfine transition frequency of the caesium-133 atom | | | | | | | |
| m | $c$: speed of light in vacuum | | X | | | | | |
| A | $e$: elementary charge | | X | | | | | |
| kg | $h$: Planck constant | | X | X | | | | |
| K | $k_B$: Boltzmann constant | | X | X | | X | | |
| Cd | $K_{cd}$: luminous efficacy of monochromatic radiation of frequency 540 × 10$^{12}$ Hz | | X | X | | X | | |

**Table 1:** *Dependencies of the defining constants on other SI base unit*

The evolution from astronomy to quantum physics in 1967 was associated with a deep conceptual change for the type of measured quantity underlying the *mise en pratique* of the definition. In astronomy, it was the angle/phase linked to the considered Earth motion that was determined theoretically as a given function of time. With quantum physics, the realization of the definition is now based on frequency measurements, with the assumption provided by the Standard Model that the atomic resonance frequencies are universal and constant, both in time and in space [5, 6, 7].

Today, the primary representation of the SI second is realized by caesium primary frequency standards, with relative frequency uncertainties at the $10^{-16}$ level offered by cold atom fountains (see https://www.bipm.org/en/time-ftp/circular-t and [8]).

Secondary representations of the SI second (SRS) are provided by rubidium or optical frequency standards (OFS). The list of recommended values of standard frequencies for transitions that may be used as SRS is regularly updated [3, 9, 10].

## 3. Main needs in TF metrology and stimulus for a new definition

With the SI second underlying the realization of other SI units, its redefinition may potentially impact a very wide range of communities. Here we consider the impact and the drive for a new definition of the SI second on the metrological community represented by the National Metrology Institutes (NMIs) and the Designated Institutes (DIs), and on the wider timing community. In addition, the findings of the CCTF survey are summarized.

### 3.1. Significance of the redefinition for the NMIs and DIs

The NMIs and DIs, as part of their mandates, strive to develop the best realizations of the SI units and build the highest accuracy primary standards. They also typically have the most demanding requirements for accessing accurate time and frequency signals because they provide the highest tier SI dissemination services for their respective countries. The current primary frequency standards have now been surpassed in terms of stability and systematic uncertainty by optical frequency standards, and, therefore, the NMIs and DIs are expected to drive the transition to the new state-of-the-art definition.

The implementation of a new definition of the SI second, based on optical standards, and an improved Coordinated Universal Time (UTC) will require the metrology labs to acquire new systems and adopt new methods. The stakeholder survey that was conducted in December 2020 to January 2021 showed an overall positive response to the redefinition plans, which indicates high levels of commitment and technical maturity that is essential to support the redefinition work.

### 3.2. Significance of the redefinition for the wider timing community

Although relatively unknown to the general public, sub-µs timing and synchronization capability has become an essential and crucial feature of most critical infrastructure, including telecommunications, energy, finance, cloud computing, transportation and space activities. Even though these applications do not require the accuracies of the optical clocks today, they, in general, depend on TF metrology.

In addition, many scientific applications require nanosecond levels of stability and/or accuracy such as radio astronomy, particle physics experiments, and time metrology. In the next five to ten years, the need for higher precision in both time and frequency is estimated to grow across all fields.

Initially, scientific applications will benefit more than industrial ones from the redefinition of the second and the development in the time and frequency metrology that this may underpin: for example, quantum communications, with some time accuracy and stability requirements at the level of femtoseconds, which is hardly achievable with current technologies.

### 3.3. Meeting current and future stakeholder needs

From the CCTF survey and other references [11-14], timing accuracy needs are currently in the range from 1 µs down to 10 ns, while future needs seem to focus below 100 ns for most users. Some scientific users highlighted the need for a sub-nanosecond timing accuracy. The most stringent fractional frequency accuracy needs are currently around 1E-14, while future needs are specified up to 1E-15 or 1E-18 for some specific users.

The most fundamental of the existing scientific applications that will be improved by a redefinition and the resulting improvement in timing infrastructure, are tests of fundamental physics, for which the levels of accuracy achievable with optical clocks can underpin tests of fundamental physical theories, including the investigation of physics beyond the standard model and time variation of the fundamental constants, the search for dark matter, gravitational wave detection, and more [15].

Better clocks will also enable higher-precision atomic and molecular spectroscopy as well as improved time synchronization for high-resolution telescope arrays and future VLBI generations [16], geopotential monitoring with centimetre resolution [17], quantum networks for quantum encrypted communications [18], and others.

These emerging fields of research that already require better TF accuracy or stability than is available today and applications that promise to transition from the research lab into commercial use in the next decades will benefit from the improved accuracy enabled by a redefinition.

A redefinition of the SI second will also lead to timing infrastructure improvements, including improved time scales and frequency transfer methods. These improvements will benefit the wider stakeholder community, including clock and equipment manufacturers and users. The redefinition of the second constitutes a required step in stabilizing and directing the technology development, standardization and adoption.

Table 2 lists the stakeholder requests for their future needs in the accuracy of frequency references. It is clear from the high level of interest in more accurate frequency reference signals that many research opportunities will arise with better access to optical clocks and better dissemination methods.

| uncertainty level | Application opportunity |
|---|---|
| 1E-14 | holdover |
| 1E-15 | spectroscopy/dark matter/secure com/holdover |
| 1E-16 | cosmology |
| 1E-17 | dark matter/connected interferometry |
| 1E-18 | positioning/real time geodesy/new clocks |
| 1E-19 | geodynamics |
| 1E-20 | relativistic geodesy/alternative theories of gravitation |

**Table 2:** *Stakeholder responses to the question: What level of frequency uncertainty would you like to access in the future?*

## 4. Options for the redefinition of the SI second

The current definition of the SI units is established in terms of a set of seven defining constants with fixed numerical values, as declared in Resolution 1 of the 26th meeting of the CGPM (2018) [19].

Three of these defining constants: $c$, $h$, and $e$, are directly embodied in the fundamental theoretical framework of general relativity and the standard model of particle physics. The defining constant for the unit of time, $\Delta\nu_{Cs}$, is a property of the Cs atom and consequentially a natural constant. The other three defining constants have a less direct connection to the fundamental framework, $k_B$, $N_A$ being conversion factors, and $K_{cd}$ being linked to the sensitivity of the human eye.

There are three options for the redefinition of the second, which all keep the same principle of applying seven defining constants but would replace $\Delta\nu_{Cs}$ by a different constant.

Option 1 consists of choosing one single atomic transition in lieu of the Cs hyperfine transition and to fix the numerical value of the frequency of this transition $\nu_{Xy}$

$\nu_{Xy} = N$ Hz, where $N$ is the defining value.

Option 2 consists of creating a defining constant based on several transitions rather than just a single one, as described in [20]. The quantity whose numerical value is used in the definition is a weighted geometrical mean of the frequency of an ensemble of chosen transitions. The unit of time is set by the relation:

$\prod_i \nu_i^{w_i} = N$ Hz, where $w_i$ and $N$ are the defining values, with the sum of all $w_i$ being equal to 1.

Option 3 consists in fixing the numerical value of one more fundamental constant, in addition to $c$, $h$ and $e$. From the fundamental standpoint, a good choice for this constant is the electron mass $m_e$ (see e.g. [21]), in which case the system of units is set by the relations:

$m_e = M$ kg,

where $M$ is the defining value, completed by the other defining relations for $c$, $h$, $e$, $k_B$, $N_A$ and $K_{cd}$.

In this system, one can see that the Compton frequency $\nu_e$ defined by $h\nu_e = m_e c^2$ has a defined value, which shows how such a system defines the unit of time. Another choice is to directly fix the numerical value of $\nu_e$ instead of $m_e$. A third choice is to fix the numerical value of the Rydberg frequency $R_\infty$ which is also linked to the electron mass via the relation $R_\infty = \alpha^2 \nu_e/2$, where $\alpha$ is the fine-structure constant. The two first choices are two different formulations for systems of units that are physically identical. The third choice defines a physically different system of units since $\alpha$ is a dimensionless constant that can only be measured and cannot be fixed by our choice.

While all three options concern primarily the definition of the SI second, they would have a formal impact on the definitions of all other base units with the exception of the mole, because these make use of the definition of the second via $\Delta\nu_{Cs}$.

To complement these formal aspects of the redefinition options, several points are worth noting. Regarding Option 1, it is anticipated that besides the primary transition selected for the definition, other transitions will contribute to realizations and disseminations of the unit of time according to the mechanism of SRS that is already in place and will be described in more detail in section 6. As a possibility associated to Option 2, it is also proposed that future revisions of the defining values $w_i$ and $N$ could be adopted by the CIPM, based on the recommendation of the CCTF and CCU, and according to a set of rules adopted beforehand by the CGPM. Rules include a quantitative criterion to trigger a revision that ensures the convergence through successive updates (see [20]). Rules are designed to ensure that revisions are made only when significant improvement of the realization and dissemination will ensue. This dynamic option is referred to as option 2b, while the option 2 with fixed values of weights and $N$ is named option 2a.

Regarding Option 2, the realization makes use of best estimates of optical frequency ratios established via the fitting procedures that are already in place of the CCL-CCTF WGFS [22]. Given these ratios, one single frequency standard based on either of the chosen transitions can realize the unit of time [20]. In addition to the conceptual aspect, i.e. the possibility to define the unit of time and the system of units using several transitions, Option 2 gives a possible approach to cope with the present context where many different atomic transitions give optical frequency standards with uncertainties near $10^{-18}$ and where the field will remain highly dynamic.

Under Option 3, the numerical value of the defining constant for the unit of time relies on experiments that presently lead to the determination of the chosen constant. The evaluations of relevant experiments are the work of CODATA and are reported in [23]. Currently, the value of $m_e$ has an uncertainty of 3.0 parts in $10^{10}$, while the uncertainty in the Rydberg constant is 1.9 part in $10^{12}$. These uncertainties are several orders of magnitude larger than the present realizations of the unit of time of the current SI system (few parts in $10^{16}$) and even further away from the capabilities of optical frequency standards ($10^{-18}$ or better). Consequently, Option 3 is not practical in the current state of science and technology.

It is also worth noting that measurements between the optical frequency domain and the current best realizations of the SI second are already done with low enough uncertainty (near $10^{-16}$, the limit of fountain frequency standards) and with sufficient redundancy to ensure the continuity between the current definition and any definition based on optical transitions.

To summarize the trade-offs between the three options, we present here their most significant respective strengths, weaknesses, opportunities, and threats in tabular form (i.e., a SWOT analysis) (

Table **3**). We note that these considerations have taken into account the needs of both the user and research communities, as assessed by the CCTF Task force via input from user surveys and BIPM workshops.

|  | Option 1 | Option 2 | Option 3 |
|---|---|---|---|
| Strengths | Offers two orders of magnitude improvement of the existing definition with significant improvement likely in the future<br><br>Maintains continuity with the current Cs definition<br><br>Intuitive extension of the existing definition<br><br>Familiar and practical, using primary and secondary realizations as we do today<br><br>The unit of time can be realized without additional uncertainty | Offers two orders of magnitude improvement of the existing definition with significant improvement likely in the future<br><br>Maintains continuity with the current Cs definition<br><br>Flexible scheme that is well matched to the current experimental situation and could adapt well to rapid progress in optical standards<br><br>Could more easily lead to a consensus on the chosen species. | Consistent with the approach adopted by CIPM based on the physical constants, $c$, $h$, $e$, and $k_B$<br><br>Direct connection to the theoretical framework of fundamental physics |
| Weaknesses | With no clear preferred transition at present, it may be hard to reach a consensus | Can be difficult to understand and convey to general users<br><br>The unit of time may be hard to realize by a single institute in isolation<br><br>The version which allows for revisions of the defining values $w_i$ and $N$ constitutes a conceptual deviation from the principle of applying fixed defining constants for the SI units as implemented in 2019.<br><br>A better uncertainty obtained with one transition alone is not enough to have a better realization of the unit<br><br>The defining constant has no physical meaning – all realizations are secondary representations<br><br>A more complex definition of time may present legal issues for some countries | Would lead to poor accuracy for time realization in the present and foreseeable future<br><br>Would represent a step backwards in time realization by four orders of magnitude (six relative to Options 1 and 2)<br><br>Would not allow continuity with the current Cs definition, which allows a much better accuracy in the realization |
| Opportunities | The many benefits associated with an improvement of a factor of 100 (or more) in the definition of the unit of time<br><br>A clear path forward for development of primary standards<br><br>Provides a stimulus for the development of commercial standards | The many benefits associated with an improvement of a factor of 100 (or more) in the definition of the unit of time<br><br>Provides a strong stimulus to explore new frequency standard options | This approach would lead to a consistent set of SI definitions that is close to the theoretical foundations of physics.<br><br>Could stimulate further research in simple atoms, calculable quantum systems and the measurements of fundamental constants |
| Threats | Depending on the quality of future OFS reports for TAI calibration, it might be difficult to provide at least as good uncertainty of dTAI after the redefinition<br><br>The new definition might rapidly become obsolete – SRS could end up dominating contributions to TAI<br><br>Could discourage future progress on frequency standards, by biasing work towards the chosen transition | Depending on the quality of future OFS reports for TAI calibration, it might be difficult to provide at least as good uncertainty of dTAI after the redefinition<br><br>A multi-species definition might lead to difficulty for industry (and NMIs) in choosing which standard to develop | There would be a severe degradation in the realization of the SI unit of time<br><br>Such a definition would break the metrological principle that redefinitions should be consistent with previous definitions within the uncertainty with which the old definition was realized |

**Table 3**: *Collection of Strengths, weaknesses, opportunities and threats of the 3 options for the redefinition, based on input from a community survey in 2022*

## 5. Criteria and conditions for the redefinition

In order to choose the best new definition and its implementation timeline, and to provide the CGPM with all the required information for making its decision, criteria and conditions (Table 4) have been defined to assure that the redefinition:

- offers an improvement by 10 to 100 of the realization of the new definition in the short term after the redefinition (reaching $10^{-17}$ to $10^{-18}$ relative frequency uncertainty) and potentially a larger improvement in the longer term (*criteria I.1, I.2, III.1 and condition III.3*), requiring the capability to compare OFS with an adequate uncertainty to validate OFS uncertainty budgets (*criteria II.1, II.2*);
- ensures continuity with the current definition based on caesium (*criterion I.3*);
- ensures continuity and sustainability of the availability of the new SI second through TAI/UTC and enables a significant improvement of the quality of TAI and UTC(*k*) as soon as the definition is changed (*criterion I.4 and conditions I.6, III.3*), relying on the reliability of OFS and TF transfer infrastructures (*conditions I.5, II.3*);
- is acceptable to all NMIs and stakeholders and enables the dissemination of the unit to broad categories of users (*criterion III.2 and conditions III.4, III.5*);

| | Mandatory criteria | Ancillary conditions | Criteria and conditions |
|---|---|---|---|
| **Frequency standards, including the contribution of OFS to time scales** | X | | I.1 - Accuracy budgets of optical frequency standards |
| | X | | I.2 - Validation of Optical Frequency Standard accuracy budgets – Frequency ratios |
| | X | | I.3 - Continuity with the definition based on Cs |
| | X | | I.4 - Regular contributions of optical frequency standards to TAI (as secondary representations of the second) |
| | | X | I.5 - High reliability of OFS |
| | | X | I.6 - Regular contributions of optical frequency standards to UTC(*k*) |
| **TF links for comparison or dissemination** | X | | II.1 – Availability of sustainable techniques for Optical Frequency Standards comparisons |
| | X | | II.2 – Knowledge of the local geopotential with an adequate uncertainty level |
| | | X | II.3 – High reliability of ultra-high stability TF links |
| **Acceptability of the new definition** | X | | III.1 - Definition allowing more accurate realizations in the future |
| | X | | III.2 – Access to the realization of the new definition |
| | | X | III-3 - Continuous improvement of the realization and of time scales after redefinition |
| | | X | III.4 - Availability of commercial optical frequency standards |
| | | X | III.5 - Improved quality of the dissemination towards users |

**Table 4:** *Mandatory criteria and ancillary conditions to ensure the benefit and the acceptability of a new definition.*

Criteria and conditions are distinguished in the following way:

- the mandatory criteria that must be achieved before changing the definition;
- the ancillary conditions that are not required to be fully achieved to change the definition but are important to ensure the best realization and exploitation of the new definition in the short and long terms. Thus, these conditions correspond to essential work that must have started before the redefinition, with a reasonable amount of progress at the time of redefinition and a commitment of stakeholders to continue their efforts on the associated activities.

Fulfilment indexes have been defined to evaluate the fulfilment level for mandatory criteria to quantitatively follow the improvements, to be aware of the remaining work to fulfil all mandatory criteria and ultimately, to decide it is time to change the definition. The details of criteria and conditions and their current fulfilment levels or progress statuses are presented in Section 8.

## 6. Optical frequency standards - Categories and characteristics

### 6.1. Types, characteristics and performance of optical frequency standards

Due to their demonstrated potential for low fractional frequency instabilities and uncertainties, there is currently considerable research activity directed towards investigating optical transitions to serve as frequency standards. These standards fall into two categories distinguished by the charge state of the atom and the method used for trapping: trapped ion optical clocks and optical lattice clocks with neutral atoms. Presently, ten optical transitions and one microwave transition ($^{87}$Rb) are recommended as SRS, as listed in Table 5. We note that due to the lower uncertainties associated with most of the optical standards themselves, the uncertainties for the realizations of the second with these standards as listed in the Table are largely determined by the uncertainty of microwave standards based on the Cs transition that enters into the recommended frequencies.

Advances in several key technologies have been critical to the rapid improvement in optical standards. To achieve a low instability, it is necessary to start with an extremely narrow linewidth clock laser. Thus, pre-stabilization of the clock laser to a high-performance optical cavity is a standard component of any high-performance standard. Fractional frequency instabilities as low as $8 \times 10^{-17}$ on 1 s timescales have been achieved with a clock laser locked to the resonance frequency of a room temperature 48 cm ULE FP cavity [24], while locking to cryogenic single-crystal optical cavities has led to frequency instabilities in the low $10^{-17}$ range [25, 26]. In addition, the development of optical frequency combs (OFC) [27, 28], which are needed to link optical frequencies directly with microwave frequencies, has made high-fidelity measurements of absolute optical frequencies at the low $10^{-16}$ uncertainty level of fountain clocks feasible. In fact, simultaneous measurements of the same optical frequency ratio with two independent OFCs have shown agreement at the level of $10^{-21}$ [29], thereby confirming the capability of OFCs to support optical frequency ratio measurements at the limit of the uncertainties of current optical clocks. These capabilities have enabled more precise (and more rapid) comparisons between standards, with many of the optical standards realizing SRS as listed in Table 5 reaching Type B uncertainties below $10^{-17}$.

The current record for systematic uncertainty of an atomic clock is held by the $^{27}$Al$^+$ quantum logic clock, with a fractional frequency systematic uncertainty of $9.4 \times 10^{-19}$ [30]. This level of performance is closely followed by that of an Yb optical lattice clock ($1.4 \times 10^{-18}$ [31]), a Sr optical lattice clock ($2.0 \times 10^{-18}$ [32]), an $^{171}$Yb$^+$ ion clock operated on the octupole (E3) transition ($2.7 \times 10^{-18}$ [33, 34]), and recently a $^{40}$Ca$^+$ ion clock ($3.0 \times 10^{-18}$ [35]). Interestingly, it seems there is not a fundamental limitation for the accuracy of the optical clocks that are being developed based on different ion and neutral atom species. Most of the currently proposed optical transitions can potentially achieve an uncertainty level below $10^{-18}$. We note that the lowest instabilities achieved at 1 s averaging time have been observed with optical lattice clocks: $4.8 \times 10^{-17}$ [32] and $6 \times 10^{-17}$ [36]. For single ion clocks, the lowest reported instabilities at 1 s are typically around $1 \times 10^{-15}$ [30, 37].

| Transition | Approximate wavelength | Recommended frequency (Hz) | Recommended relative uncertainty | Used to calibrate TAI scale interval |
|---|---|---|---|---|
| $^{199}$Hg | 265 nm | 1 128 575 290 808 154.32 | 2.4E-16 | |
| $^{27}$Al$^+$ | 267 nm | 1 121 015 393 207 859.16 | 1.9E-16 | |
| $^{199}$Hg$^+$ | 282 nm | 1 064 721 609 899 146.96 | 2.2E-16 | |
| $^{171}$Yb$^+$(E2) | 436 nm | 688 358 979 309 308.24 | 2.0E-16 | |
| $^{171}$Yb$^+$(E3) | 467 nm | 642 121 496 772 645.12 | 1.9E-16 | |
| $^{171}$Yb | 578 nm | 518 295 836 590 863.63 | 1.9E-16 | Yes (4 institutes) |
| $^{88}$Sr$^+$ | 674 nm | 444 779 044 095 486.3 | 1.3E-15 | |
| $^{88}$Sr | 698 nm | 429 228 066 418 007.01 | 2.0E-16 | |
| $^{87}$Sr | 698 nm | 429 228 004 229 872.99 | 1.9E-16 | Yes (3 institutes) |
| $^{40}$Ca$^+$ | 729 nm | 411 042 129 776 400.4 | 1.8E-15 | |
| $^{87}$Rb | | 6 834 682 610.9043126 | 3.4E-16 | Yes (1 institute) |

**Table 5:** *List of secondary representations of the second adopted by the 22nd CCTF (March 2021) [9].*

### 6.2. Ratio measurements between frequency standards

In order to verify the predicted levels of performance for these standards, there has been a great effort over the past decade to perform measurements of frequency ratios between co-located or remotely located standards. Such comparisons can be based on the same transition or different transitions. Comparing different optical standards based on the same transition provides a way to validate uncertainties by verifying that the realized transition frequencies agree within stated uncertainties. To date, several such comparisons performed within the same institute have reached an overall uncertainty better than $5 \times 10^{-18}$ [31, 34], with the lowest reaching $1 \times 10^{-18}$ [31]. Comparisons between standards based on the same transitions from different institutes are at the level of $5 \times 10^{-17}$ [38]. We note that comparisons between clocks in different locations are much more challenging because they involve either remote comparison, which can be limited by the instability of long-distance time transfer capabilities or transportable standards, which generally have lower levels of performance than their lab-based counterparts. In general, such comparisons are of utmost importance to validate the frequency standards' uncertainties.

Equally valuable are frequency ratios measured between standards based on different transitions. Such ratios between unperturbed atomic transitions are significant, because they are dimensionless quantities given by nature. As a result, two independent measurements of such ratios should coincide within the combined measurement uncertainties. Thus, comparisons between independent measurements of given ratios provide further means to validate stated uncertainties of optical frequency standards. We note that such measurements almost always rely on optical frequency combs to span the frequency gap between standards. Therefore, comparing independent measurements of a given optical frequency ratio tests not only the stated uncertainties of optical standards themselves, but those of the combs (and any other optical frequency metrology capabilities relevant to the use of optical frequency standards). To date, the most accurate measurement of an optical frequency ratio has a fractional uncertainty of $6 \times 10^{-18}$ (between two labs about 2 km apart) [38, 39]. A few optical ratios have been measured multiple times by different institutes, thereby enabling first comparisons of such measurements at uncertainty levels ranging from $3 \times 10^{-17}$ to $2 \times 10^{-16}$.

We also emphasize that frequency ratio measurements between optical and microwave standards are common and serve to validate our capabilities to connect the optical domain with the microwave domain, as well as to link a potential future definition to the current one. The accuracies of such measurements are now at the limit of the primary standards based on Cs ($\sim 10^{-16}$). In the last few years, many such absolute measurements of optical standards have been performed by comparison with TAI, whose rate with respect to the SI second is provided by BIPM publications, based on the currently available reports from primary and secondary frequency standards (https://www.bipm.org/en/time-ftp/circular-t). Several groups have performed extended measurement campaigns involving both optical and microwave clocks that have lasted from several months [40, 41, 42, 43, 44] to several years [45, 46, 47]. Although not continuous, these campaigns were realized by performing multiple measurements over a given time span.

Taken as a whole, the resulting ensemble of high accuracy measurements of atomic frequency ratios published after peer-review provides an overdetermined dataset from which one can determine the best values for these atomic frequency ratios, using an adjustment procedure. This task is done on a regular basis by the CCL-CCTF working group on frequency standards (CCL-CCTF WGFS). The resulting output of this calculation provides the basis for the recommended values and uncertainties of frequency standards shown in Table 5 [9]. In addition, given the strongly overdetermined nature of the dataset, this adjustment provides a global validation of the status of high accuracy atomic frequency standards and of related measurement capabilities, as described in [3]. In the last implementation reported to the 22nd meeting of the CCTF on 19 March 2021, the adjustment took into account 105 measurements (69 in 2017), including 33 optical frequency ratios (11 in 2017) and 72 absolute frequency measurements (58 in 2017). We note that it is necessary to take into account correlations (483 for the latest adjustment) between these measurements to perform the calculation correctly [10].

### 6.3. Ongoing research activities and future prospects for optical standards (new transitions, improved stability, transportable standards)

Despite the considerable progress to date in optical clock performance, there remains much room for further improvements in terms of clock stability, uncertainty, and robustness. Reduced clock instability is not only useful in direct timing applications, but the extremely low uncertainty of optical clocks is only useful if the statistical uncertainty (Allan deviation) can be reduced to the evaluated uncertainty level at a practical averaging time for the measurement application. Improvements in the observed stability of optical lattice clocks and long-lived ion transitions ($^{27}$Al$^+$, $^{171}$Yb$^+$ (E3)) are ongoing but are technically challenging, as they require ultra-stable lasers with coherence times of several seconds to minutes. In addition to continued advances in cavity performance mentioned earlier, there are efforts in parallel to develop novel measurement protocols that mitigate the limitations caused by reference cavity noise, such as zero-dead time interrogation [36], correlation spectroscopy [48, 49], and dynamic decoupling of laser phase noise in compound atomic clocks [50]. It is anticipated that the use of compound clocks could improve the stability of single ion clocks with long clock transition lifetimes to levels comparable to that of optical lattice clocks [50]. For ion species with shorter lifetimes, the stability can be improved directly by increasing the number of ions, but this approach requires special care in the selection of the atomic transition and the control of the systematic shifts to preserve accuracy [51,52]. Entanglement in multi-ion or neutral atom clocks offers the potential for a stability beyond the standard quantum limit and thus could be a method to further improve the stability of optical clocks [53]. A new type of clock with high relative stability has been demonstrated recently, called a "tweezer array optical clock" that balances the benefits of non-interacting particles as found in single-ion clocks with the large number of atoms as found in optical lattice clocks [54].

Another critical aspect for the spread of optical clock performance throughout the clock community will be the demonstration of high duty cycle, high performance, robust optical systems. In this direction there has been considerable effort with many systems under development. Indeed, all major subsystems of an optical clock with laser cooled atoms or ions have already been developed as robust transportable devices for autonomous operation, which have been partially tested for operation in space. This includes vacuum systems and traps for atoms [55] and ions [56], tunable laser systems for cooling and interrogation, optical reference cavities for obtaining a narrow linewidth of the reference laser [57, 58, 59], and optical frequency combs for transfer of the optical stability to a microwave output signal [60]. However, the integration of an optical clock from the subsystems also requires the robust optical alignment of multiple laser beams and the monitoring, control and adjustment of a few dozen electrical and mechanical parameters. Fully integrated prototype systems that have been used as transportable optical clocks on the footprint of a small trailer have been demonstrated for a Sr optical lattice clock [61, 91] and for a clock with a single trapped $Ca^+$ ion [62].

Some groups have demonstrated high clock operation uptimes, for example 80.3 % for a duration of six months [41], 93.8 % uptime for a period of 10 days [44]. More recently fully autonomous operation for two weeks with 99.8 % uptime at $2 \times 10^{-17}$ systematic uncertainty inside a laboratory has been demonstrated for the OptiClock based on the E2 transition of $^{171}Yb^+$ [63, 64]. The system fits inside the volume of two 19-inch racks and has been developed by PTB jointly with industry [64]. Optical clocks with (nearly) 100 % uptimes for one month of continuous operation or longer are expected to become common in the next few years. These results indicate that the development of a turn-key autonomous optical clock is technically feasible at a performance level that is superior to available microwave frequency standards and shows the way towards a commercial high-performance optical reference.

Finally, one of the most exciting directions in optical clock research today is the search for transitions that have still lower sensitivities to external fields than current optical clocks in an effort to further reduce clock uncertainties. Some of these include a nuclear transition in $^{229m}Th$ [65], and transitions in highly-charged ions [66, 67] and lutetium ions [52]. While all of these systems present their own technological challenges, they could well be among the main candidates for future optical clocks with performance at the 19th and 20th digits.

## 7. TF transfer and time scales – Categories and characteristics

### 7.1. TF transfer

Remote comparison of time scales and frequency standards is possible using various space-based microwave techniques for time and frequency transfer, including Global Navigation Satellite Systems (GNSS), Two-way Satellite Time and Frequency Transfer (TWSTFT), and Very Long Baseline Interferometry (VLBI) radio antennas. In the last decade, optical techniques using fibre optic links have offered greatly improved stability and accuracy. Innovative satellite transfer in the optical domain is also envisaged. Lastly, Transportable Optical frequency standards or Clocks (TOCs) used as travelling standards can support a redefinition of the second that requires comparisons at an accuracy level of $10^{-18}$ and global geographical coverage.

GNSS time transfer is a one-way technique used since the 1980s, notably for the realization of UTC. A collaboration with the International GNSS Service (IGS) has led to the use of Precise Point Positioning (PPP) for time and frequency comparisons and development of the integer ambiguity PPP technique (IPPP), which to date offers the best long-term stability among the GNSS techniques. Fig. 1 [68] shows that IPPP provides time transfer with a modified Allan deviation of $7 \times 10^{-16}/\tau$, where $\tau$ is the duration in days of continuous phase measurements. A twofold improvement is expected using satellites from all the GNSS, as opposed to just GPS as at present.

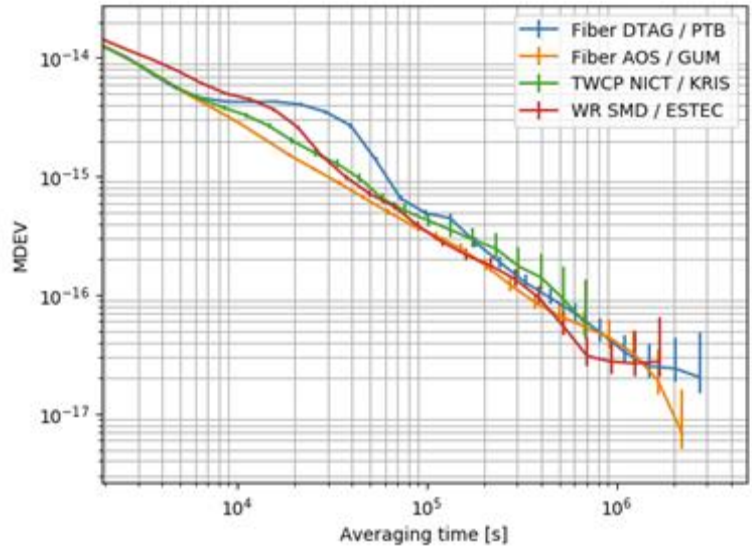

**Figure 1:** *Modified Allan deviation of the comparison between IPPP and several other high accuracy techniques: The optical fiber links DTAG-PTB (blue), AOS-GUM (orange) and SMD-ESTEC (red) and the two-way carrier phase link NICT-KRIS (green) [68]*

TWSTFT, the second intercontinental-capable satellite-based microwave method, typically employs the code-phase of a signal modulated by a pseudorandom noise code sent and received by microwave link via a geostationary telecommunications satellite, at Ku-band frequencies [69]. Improved performance is achieved by the use of Two-Way Carrier-Phase (TWCP), which exploits carrier-phase measurements, with an instability of a few parts in $10^{16}$ at one day. Further results [70] indicate that TWCP performs at least as well as IPPP in terms of stability. Fig. 2 shows the modified Allan deviation of Code Phase and Carrier Phase TWCP.

In addition, a recently implemented software-defined receiver (SDR) successfully reduced the long-term instability by about a quarter [71]. Similar technology is expected to be applied to the transmitters for further improvement resulting in integrated digital modems that are an important step to improve TWSTFT beyond the current state of art. Moreover, in order to reach to the sub 1e-17 level it is essential to improve on modeling of all non-reciprocal error sources, such as signal propagation, atmospheric turbulence, and relativistic effects [72].

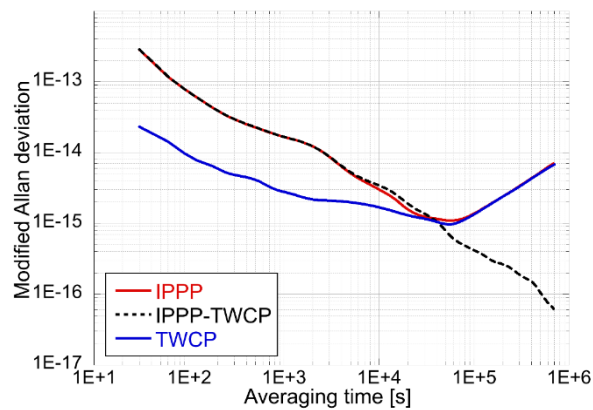

**Figure 2:** *Modified Allan deviation of UTC(NICT)-UTC(KRIS) from MJD 57851 to 57883 measured by different techniques [70]*

VLBI utilizes the reception of radio signals from extragalactic radio sources, with the time difference between the arrivals of the signals measured at two antennas equipped with local atomic clocks. Using VLBI, the frequency of an Yb and a Sr optical standard has been compared [73], with a statistical uncertainty from the VLBI link of $9 \times 10^{-17}$ over 300 hours of measurements.

Using optical communication, satellite-based comparisons were demonstrated with the Time Transfer by Laser Link (T2L2), onboard the Jason-2 satellite [74]. Three T2L2 links were compared with IPPP links [75], with the standard deviation of the time difference well below 100 ps. Promising results have also been obtained using terrestrial free-space optical time and frequency transfer, using cw or coherent pulsed lasers. For both, uncertainties of parts in $10^{16}$ in a few minutes have been achieved over distances up to tens of kilometers. The synchronization of two clocks 28 km apart below 1 fs within 100 s, even at high Doppler velocities of up to ±24 m/s, and under stable weather conditions has been shown [76]. A comparison at 113 km with modified Allan deviation of $10^{-19}$ at $10^4$ seconds was also reported [77], the first evidence of the method compatibility with Low Earth Orbit satellites. Figure 3 indicates both results.

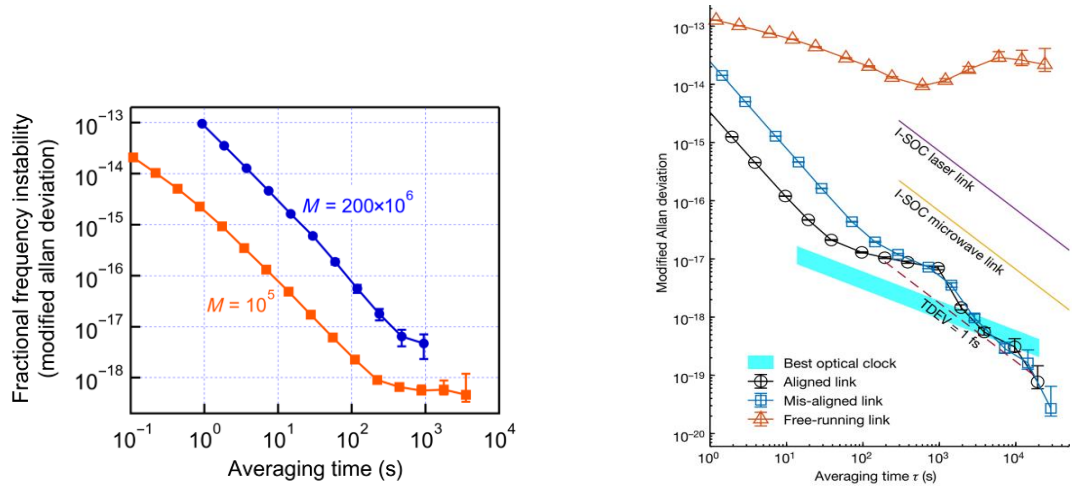

**Figure 3**: *Free space optical link fractional frequency instability. Left: Modified Allan deviation over 28 km [76]. M is the ratio $f_r/\Delta f_r$, where $f_r$ is the nominal repetition rate, and $\Delta f_r$ is the real difference between the repetition rates of the two involved combs. Right: Modified Allan deviation over 113 km [77] (Black circles, well-aligned free-space time–frequency link; blue squares, mis-aligned link; orange triangles, free-running link). The performances of the best optical clock, the I-SOC (Space Optical Clock on the International Space Station) laser link, the I-SOC microwave link and the TDEV of 1 fs are also shown.*

Optical fibres offer several key advantages compared to free-space techniques: high isolation from external interference; high bandwidth; and low propagation losses, when compensated by optical amplifiers and regeneration devices, at distances more than 1000 km. For time and frequency comparisons, three main methods are used: CW light from an ultra-stable laser, without modulation; modulated laser light (amplitude, frequency, or phase modulation); and protocol-based signals, based on digital data transfer.

Propagation of optical signals in optical fibres for frequency comparisons offers two main choices: bi-directional fibre links, providing the best performance, and unidirectional fibre links, which are easier to implement on common telecommunication networks. Submarine links are less noisy than terrestrial links [78], as shown in Figure 4.

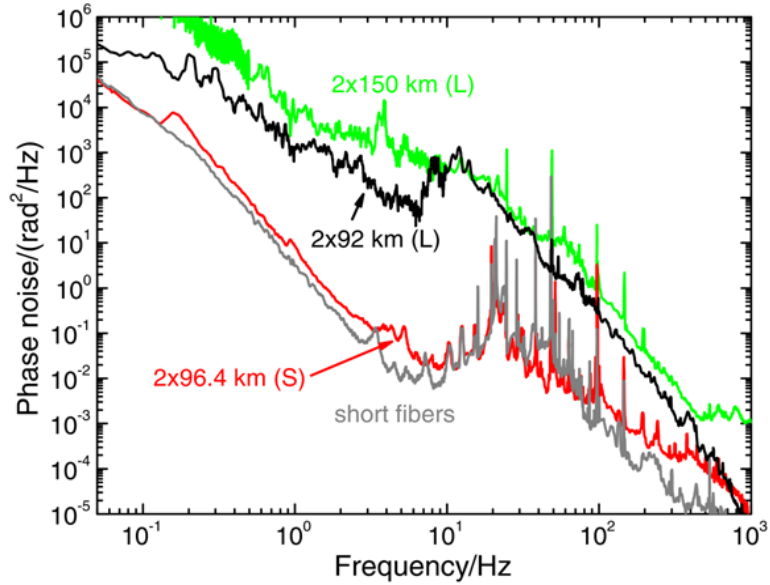

**Figure 4**: *Submarine testbeds, round-trip phase noise [78]. L indicates land links, S submarine links. Red line: submarine 2 × 96.4 km link; grey line: measurement noise floor; green line: 2 × 150 km fibre along highway; black line: 2 × 92 km fibre along highway (other area).*

Optical frequency transfer over fully bi-directional links [79] exhibits typical Allan deviations of ~$10^{-15}$ at 1 s and <$10^{-18}$ for greater than 100 s, over 100 to 1000 km long links. There is no systematic frequency shift reported so far at the level of $10^{-18}$. Conversely, optical frequency transfer over unidirectional links has demonstrated an Allan deviation of ~$10^{-15}$ at 1 s integration time, unidirectional links has demonstrated an Allan deviation of $7 \times 10^{-17}$ for averaging times between 30 s and 200 s [80]. There is no systematic frequency shift reported so far in the range of $10^{-16}$ [81]. Modulation of the optical carrier frequency enables a frequency reference in the radio and microwave domain (10 MHz -10 GHz) to be transmitted, with typical uncertainty less than $10^{-17}$ at $10^4$ s. Latest synchronization experiments report 300 km free space link and demonstrate a sub ps capability [82].

Time transfer over fibre can be in the radio/microwave domain (10 MHz - 10 GHz) or in the optical domain. In either case, the technique requires the modulation (amplitude, phase or frequency) to be tied to a time scale. The time uncertainty is less than 1 ns, approaching tens of picoseconds, in particular with White Rabbit Precise Time Protocol [83] and the ELSTAB technique [84].

Transportable optical clocks offer the best immediate prospects to meet the criteria for the redefinition of the SI second in regard to the required accuracy level and geographical coverage. As described above, space microwave techniques need to significantly improve their uncertainty levels. Fibre techniques meet the required uncertainty, but obtaining global coverage requires a large effort and investment. Satellite-based optical comparisons have not yet been demonstrated on a full metrological and operational basis. On the other hand, several TOCs have already reported the performance results that meet the redefinition requirements. An accuracy ranging from $10^{-15}$ down to parts in $10^{18}$ has been reported for several $^{87}$Sr TOCs [85-87]. A bosonic $^{88}$Sr TOC achieved $2 \times 10^{-17}$ uncertainty [88]. TOCs based on ions have also been reported: a Ca$^+$ TOC with a systematic uncertainty of $1.3 \times 10^{-17}$ [89]; an Al$^+$ standard, with four main biases evaluated at the $10^{-18}$ level [90]; and a Yb$^+$ standard demonstrated with $10^{-17}$ accuracy [64]. In addition to their role in the redefinition, the TOCs are essential tools for chronometric levelling and some have already been used for this purpose [86,89,91].

The time and frequency transfer techniques described above allow us to compare timescales and their scale intervals around the world. We can also compare the scale intervals by evaluating them with respect to locally available accurate frequency standards. However, this assumes that we have knowledge of the geopotential at the clocks location, since the atomic clocks generate their proper time and the tick rate is affected by the relativistic frequency shift. We should also note that International Atomic Time (TAI) is defined in Resolution 2 of the 26th CGPM (2018) as a realization of Terrestrial Time, which has a reference potential of $W_0$. Thus, the local geopotential needs to be obtained with respect to $W_0$ particularly for the calibration of the TAI scale interval. For the modelling of the geopotential, satellite data only provides information valid at a spatial resolution of 200 km or worse. Combining regional information from gravity measurements with the global model as well as the results of the levelling from the nearest reference to the trapped atoms, the gravity shifts of optical clocks in some metrological laboratories are now evaluated with uncertainty at the mid $10^{-18}$ level or better [92, 93, 94].

### 7.2. Time scales

Resolution 2 of the 26th CGPM (2018) states that Coordinated Universal Time (UTC), based on TAI, is the only recommended international time reference and provides the basis of civil time in most countries. Thus, the scale interval of TAI needs to be maintained with respect to optical frequency standards (OFS) for the redefinition of the second. The future TAI should have at least a similar or better performance than the current realization of TAI, which is nowadays calibrated mainly by microwave-based primary frequency standards. To this end, more than ten days of regular operation of optical clocks or a local timescale steered by an optical clock is required, in each Circular T reporting period, since the frequency link of local clocks to TAI is made by GNSS or TWSTFT. For the determination of TAI, the BIPM employs an uncertainty of $\sim 10^{-15} / (t/5)$, where $t$ is the signal integration time in days [95].

The capabilities for TAI calibration of several specific optical frequency standards have been examined by the CCTF Working Group on Primary and Secondary Frequency Standards (CCTF-WGPSFS), and as a result, eight OFSs, recognized at present as Secondary Frequency Standards (SFS), have contributed to TAI.

The first data for TAI calibration with an OFS was obtained in 2014 by an optical $^{87}$Sr lattice clock from SYRTE [44], applied for TAI calibration in 2017, and since mid-2021, at least one SFS has calibrated TAI every month. The BIPM incorporates the data from SFS into the TAI steering with additional uncertainty $u_{\mathrm{srep}}$, which is determined by the uncertainty in the CIPM recommended frequency of the SFS (Table 5). The recent update of CIPM recommended frequencies has reduced $u_{\mathrm{srep}}$, leading to an increased total weight of typically more than 10 % for all SRS for the determination of the TAI scale interval. The stated uncertainties from the laboratories, ignoring the recommended uncertainty ($u_{\mathrm{srep}}$) of the SRS, range from $1.9 \times 10^{-16}$ to $3.3 \times 10^{-15}$, limited primarily by dead time and link uncertainties. Until now, the lowest uncertainty in the SFS data submitted to TAI was reached by the NICT-Sr1 in *Circular T* 408, and IT-Yb1 in *Circular T* 411.

The calibrations provided from all OFSs [31, 41, 44, 96, 97, 98, 99] are so far consistent with those provided by primary frequency standards (see also https://webtai.bipm.org/database/d_plot.html). The development of OFSs with high uptimes over the typical reporting intervals of 15 to 30 days, the development of better local oscillators, and advances in frequency transfer are crucial goals to obtain significant improvements in the stability of TAI.

UTC is a post-processed timescale determined by the BIPM. For civil time, time and frequency metrology laboratories generate and provide real-time signals equivalent to UTC. These real time signals are called UTC($k$), denoting a real-time UTC generated at the laboratory "$k$". In general, such a UTC($k$) is often employed as a national standard time with the addition of a time offset appropriate to the respective time zone. For the future redefinition of the second, UTC($k$) generated or at least steered by an optical clock is one ancillary condition. UTC($k$) time scales must be continuous, whereas it is unrealistic at this point to operate optical clocks completely without dead time. The operation of multiple optical clocks for redundancy is not yet realized since maintaining multiple optical clocks is a difficult task and the procedure to switch between optical clocks has not yet been studied. On the other hand, intermittent operation of an optical clock enables generation of a real-time timescale steered by the optical clock [100 - 104]. Here, the source oscillator is still a microwave oscillator (hydrogen maser), but the scale interval is tuned with respect to an optical clock. In some metrology institutes, a similar generation of UTC($k$) has already been successfully implemented for some time utilizing caesium fountain frequency standards [105 -108]. In future, an all-optical timescale is expected [109], particularly for improvement of the short-term stability. Here, a CW laser, stabilized to a stable optical cavity, would play the role of the source oscillator. Considerable progress in mode-hop free operation of CW lasers has been made in the last decade.

### 8. Fulfilment levels of mandatory criteria – progress statuses for ancillary conditions

Details on mandatory criteria and ancillary conditions are presented in Sections 8.1 to 8.3 for OFS, TF transfer and the acceptability of a new definition. A synthesis of the fulfilment levels of mandatory criteria in 2022 is shown in Figure 5. Fulfilment regions have been defined, from very low fulfilment levels (<30 %, region in red) to satisfactory fulfilment levels (90-100 % and above, region in green). The vertical dashed blue line defines the threshold above which the criteria can be considered as fulfilled.

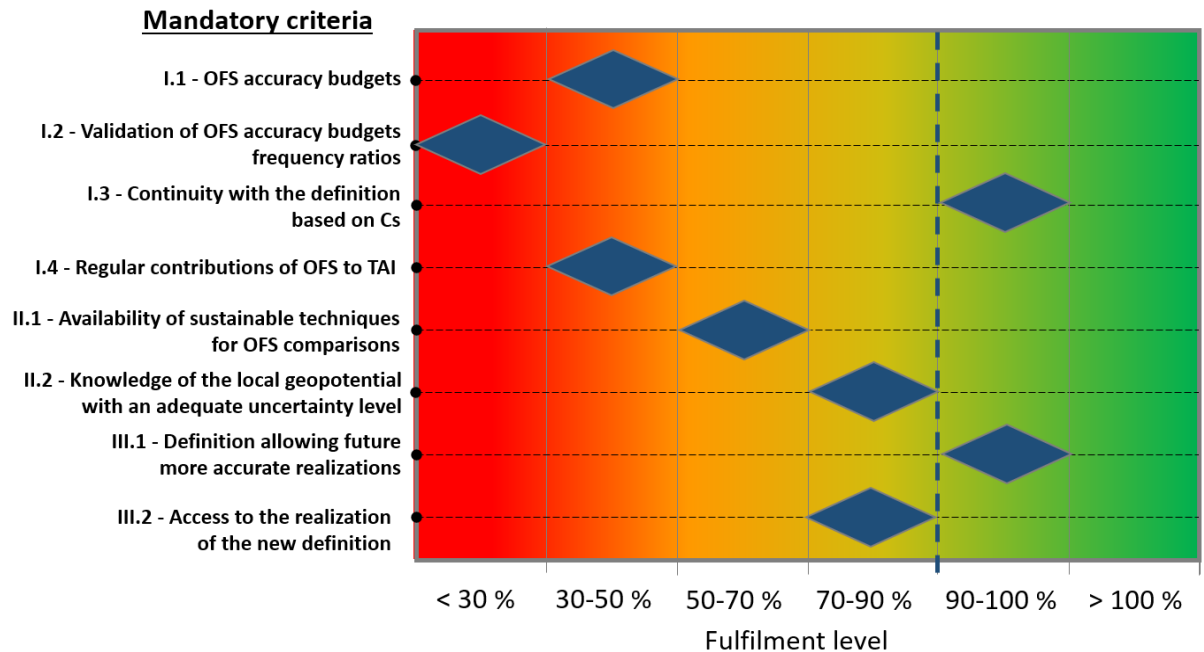

**Figure 5:** *Fulfilment levels of mandatory criteria in 2022*

While for certain criteria the fulfilment seems almost achieved, for others the fulfilment is more challenging. Due to the considerable number of optical frequency standards under development, good progress has been made on OFS performance Criteria I.1 and I.3 (fulfilment levels close to 50 % and 100 %, respectively) and on their contributions to TAI Criterion I.4 (fulfilment level of 30-50 %). Regardless of which redefinition option is chosen, the realizations of the definition will be accessible widely and their accuracy will likely continue to improve in the future with further developments on OFS (Criteria III.1 and III.2). However, the challenges associated with limited resources for developing multiple standards in one institute (along with limitations in long distance time transfer) have led to a low fulfilment level (< 30 %) for the optical frequency standards' comparison Criterion I.2.

The fulfilment of the criterion II.2 related to the knowledge of the geopotential is achieved in the majority of NMIs operating an OFS.

For the criteria II.1, a sustainable technique for OFS comparison at the proper uncertainty level is more challenging. Over intracontinental scales (baselines of about 1000 km), the requirement is fulfilled by optical fibre links, even if a significant effort for regular comparison campaigns should be addressed.

### 8.1. Criteria and conditions related to optical frequency standards and their contribution to time scales

This section contains a detailed description of the criteria and the estimation of their fulfilment level in 2022.

**Mandatory criterion I.1 - Accuracy budgets of optical frequency standards**

> I.1.a - At least three optical frequency standards based on the same reference transition, in different institutes, have demonstrated evaluated relative frequency uncertainties $\lesssim 2 \times 10^{-18}$ based on comprehensive, comparable and published accuracy budgets.

> Fulfilment level: 20-40 % [91, 31, 34]

> I.1.b - At least three frequency evaluations of optical frequency standards based on different reference transitions, either in the same institute or different institutes, have demonstrated evaluated uncertainties $\lesssim 2 \times 10^{-18}$ based on comprehensive, comparable and published accuracy budgets.

Fulfilment level: 80-100 % [30 - 32].

→ Overall Fulfilment level of criterion I.1: 30-50 %

**Mandatory criterion I.2 - Validation of optical frequency standard accuracy budgets – Frequency ratios**

I.2.a - Unit ratios (frequency comparison between standards with same clock transition): at least three measurements between OFS in different institutes in agreement with an overall uncertainty of the comparison $\Delta\nu/\nu \lesssim 5 \times 10^{-18}$ (either by transportable clocks or advanced links). Applicable to at least one radiation of I.1.

Fulfilment level: 0-20 % [91, 31, 34].

Strictly speaking the reported measurements of unit ratios are not between different institutes and should not count in this fulfilment level. Nevertheless, a fulfilment level at 0-20 % has been assigned based on these in house comparisons with uncertainties significantly lower than $5 \times 10^{-18}$ that can be considered as the first step in the right direction.

I.2.b - Non unit ratios (frequency comparison between standards with different clock transitions): at least five measurements between standards among I.1 or other, each ratio measured at least twice by different institutes in agreement with an overall uncertainty of the comparison $\Delta\nu/\nu < 5 \times 10^{-18}$ (either by direct comparisons, transportable clocks or advanced links).

Fulfilment level: 0-20 % [43]. Again, this measurement alone is not valid in terms of the criterion which demands ratio measurements « twice by independent » institutes. However, it is the first measurement at about the required uncertainty level, and it is considered the first step towards the fulfilment of this index.

→ Overall Fulfilment level for Criterion I.2: < 30 %

**Mandatory criterion I.3 - Continuity with the definition based on Cs**

There are at least three independent frequency evaluations of the optical frequency transitions utilized by the standards in I.1) with TAI or with three independent Cs primary frequency standards (in different or the same institutes), possibly via optical frequency ratio measurements, where the measurements are limited essentially by TAI or by the uncertainty of these Cs frequency standards ($\Delta\nu/\nu < 3 \times 10^{-16}$).

→ Fulfilment level: 90-100 % [44, 45, 46, 97, 98, 104, 110, 111, 112]

**Mandatory criterion I.4 - Regular contributions of optical frequency standards to TAI (as secondary representations of the second)**

At least three state-of-art calibrations of TAI (uncertainty $\lesssim 2 \times 10^{-16}$ without counting the recommended uncertainty of the secondary representation of the second $u_{\text{srep}}$) each month from a set of at least five Optical Frequency Standards for at least one year. Check that there is no degradation of TAI if its calibrations were done by OFS considered as primary standards and Cs frequency standards considered as secondary standards.

Fulfilment level: 30-50 % [see https://www.bipm.org/en/time-ftp/circular-t, and https://webtai.bipm.org/database/show_psfs.htm, https://webtai.bipm.org/database/d_plot.html]

**Ancillary condition I.5 – High reliability of OFS**

Reliable continuous operation capability of OFS, in a laboratory environment, with the appropriate level of uncertainty.

Progress status: Typical uptimes of OFS over measurement durations > 10d currently cover a wide range from a few percent to 90 % [44, 112, 113], and https://www.bipm.org/en/time-ftp/circular-t

**Ancillary condition I.6 - Regular contributions of optical frequency standards to UTC(*k*)**

Progress status : Preliminary tests of UTC(*k*) steered by an OFS [100 - 103, 109]

## 8.2. Criteria and conditions related to TF links for comparison or dissemination

**Mandatory criterion II.1 – Availability of sustainable techniques for optical frequency standard comparisons**

Availability and sustainability of transportable clocks or TF links with uncertainties $< 5 \times 10^{-18}$ for frequency comparisons between at least NMIs operating optical frequency standards of I.1), on a national / intracontinental basis (baseline up to about 1000 km). Capability of repeated uncertainty estimations of these links.

→ Fulfilment level: 50-70 %  [91, 114, 115]

**Mandatory criterion II.2 – Knowledge of the local geopotential with an adequate uncertainty level**

Knowledge of geopotential differences for NMIs operating OFS of I.2) to be consistent with the uncertainty budget of a frequency comparison between OFS using advanced links, i.e. including the uncertainty budget of the two OFS and of the link. Knowledge of local geopotential for NMIs operating OFS of I.4) with an uncertainty corresponding to a frequency uncertainty $\lesssim 10^{-17}$, for the calibration of TAI.

→ Fulfilment level: 70-90 % [86, 92, 93, 94,38] and https://www.bipm.org/en/time-ftp/data

**Ancillary condition II.3 – High reliability of ultra high stability TF links**

On-demand continuous operation capability of TF links over sufficient durations that do not limit OFS comparisons and their regular contributions to TAI.

Progress Status: a few months continuous operation of fibre links for intracontinental comparisons [114,116] but no existing link allowing OFS intercontinental comparisons without degradation.

### 8.3. Criteria and conditions related to the acceptability of the new definition

**Mandatory criterion III.1 - Definition allowing future more accurate realizations**

The new definition must be long lasting. On the short term (just after the redefinition), it must ensure an improvement by 10/100 of its realization with OFS, i.e. reaching $10^{-17}/10^{-18}$ relative frequency uncertainty. On the longer term, it must have the potential for further improvement of the realization of $10^{-18}$ and beyond in order to avoid any early obsolescence of the definition.

→ Fulfilment level: 100 % (To be confirmed, based on the chosen option for the redefinition, but no identified fundamental effect limiting OFS accuracy at $10^{-18}$ level for all species in I.1, and some newer systems have the potential to go beyond $10^{-18}$)

**Mandatory criterion III.2 - Access to the realization of the new definition**

> III.2.a Realization / "mise en pratique" of the new definition must be easily understandable with a clear uncertainty evaluation process;

> Fulfilment level: 0 % (No existing document; pending the choice of the redefinition option)

> III.2.b - Access for NMIs and high accuracy users to primary or secondary realizations of the new definition;

> Fulfilment level: 100 % (To be confirmed, based on the chosen option for the redefinition, but primary or secondary representations of the SI second will continue to be accessible via metrology institutes or TAI)

> III.2.c - Cs frequency standards ensure a secondary realization of the new definition.

> Fulfilment level: 100 % (existing TAI architecture will be maintained at current level or better and Cs will be a secondary representation of the second)

→ Overall Fulfilment level for Criterion III.2: 70-90 %

**Ancillary condition III.3 - Continuous improvement of the realization and of time scales after redefinition**

Commitment of NMIs to make the best effort to:

- improve and operate optical frequency standards that provide primary or secondary realizations of the new definition (reliable / continuous operation, regular contributions to TAI, …);

Progress status: Several OFS are already in operation and used by the CCL-CCTF Working Group on Frequency Standards (CCL-CCTF-WGFS) to calculate the Recommended values of standard frequencies 2021 [10]
- maintain the operation of Cs fountain standards over the appropriate duration;
Progress status: 12 Cs fountains in operation [117- 125]
- development of new OFS;
Progress status: Several other atomic species are being investigated as potential candidates for the next generation, for example $^{229}$Th$^+$, Lu$^+$, Cd, and several highly charged ions. The most recent references can be found for example in [Proceedings of the annual IEEE IFCS https://ieee-uffc.org/symposia/ifcs, and EFTF conferences https://www.eftf.org/]

**Ancillary condition III.4 - Availability of commercial optical frequency standards**

Progress status: No available commercial OFS

**Ancillary condition III.5 - Improved quality of the dissemination towards users**

Progress status of TF links (GNSS, TWSTFT, Fibre / Internet) for the dissemination of the definition towards users:

- Frequency stability: $10^{-17} – 10^{-16}$ for satellite microwave techniques (GNSS, TWSTFT); $10^{-20}$ level for fibre links [126]
- Time accuracy: 1 ns for satellite microwave techniques (GNSS, TWSTFT); 50 ps for fibre links [127]

## 9. Schedule, conclusions, and perspectives

The possible redefinition scenarios depend on capabilities of optical frequency standards and their envisaged evolution, considering their performance, their readiness for sustainable contributions to the realization of time scales, especially TAI, and also their potential for commercial availability, and space qualification. A roadmap also needs to address TF transfer techniques required for the comparison of atomic clocks, for the construction of international time scales, and for the dissemination of reference signals to users, with an adequate uncertainty level.

Depending on the achievements and the development progress, the CCTF envisaged the possible three schedule options for the redefinition (Figure 6).

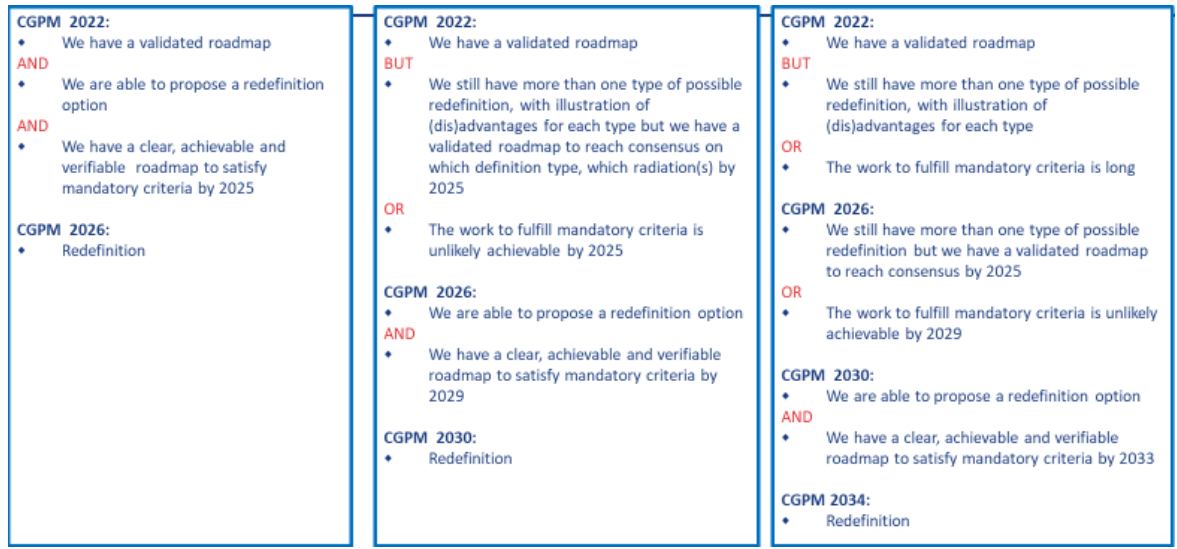

**Figure 6**: *Scenarios for the roadmap.*

It appeared clear that a redefinition at the 28th meeting of the CGPM (2026) was unrealistic since today there is no consensus on the preferred option and still some important work to do to fulfil all mandatory criteria. The 28th CGPM (2026) could validate a roadmap towards a redefinition in 2030 if, in 2026, there is a consensus on the redefinition option to be chosen and if the work to fulfil mandatory criteria is likely to be achievable by 2030. If a redefinition is not possible in 2030, it will have to be postponed until the meeting of the CGPM to be held in 2034 or the following one. But, with this third scenario, it will require the continued operation of Cs fountains primary frequency standards until the late 2030s.

The redefinition will be the occasion to further educate stakeholders on the concept of metrological traceability and the best practices for accuracy and stability measurements and their specification. The CCTF will set up a subgroup to address this particular matter and educate the public about the redefinition.

In November 2022, the 27th CGPM approved Resolution 5 [128] corresponding to the CCTF roadmap towards the redefinition of the second as presented in this paper, with a preferred scenario leading to a redefinition at the 29th CGPM (2030). This scenario is realistic, even if there is still considerable work to converge on a preferred option and to fulfil all mandatory criteria by pushing the limits of optical frequency standards and T/F transfer. All these efforts will be determining factors in reaching the goal of a new definition of the SI second with an improved quality of the *mise en pratique*, in order to serve current and future needs in metrology and to foster scientific and technological applications at the highest accuracy.

## 10. Authors contribution

This paper is based on the work of the CCTF Task Force on the "Roadmap to the redefinition of the second". The Task Force was chaired by N. Dimarcq, P.Tavella, and formed by three subgroups, whose members are listed below.

| Subgroup | chair | Executive secretary | members |
|---|---|---|---|
| A | M.Gertsvolf, G. Mileti | F. Meynadier | J. Bartholomew, P. Defraigne, E. A. Donley, P. O. Hedekvist, I. Sesia, M. Wouters |
| B | S.Bize, C.W. Oates, E. Peik | G. Petit | P. Dubé, F. Fang, T. Ido, F. Levi, J. Lodewyck, H. S. Margolis, D. Newell, S. Slyusarev, S. Weyers, J.-P. Uzan, M. Yasuda, D.-H. Yu |
| C | D.Calonico, T. Ido | G. Panfilo | P. Defraigne, E. A. Donley, M. Fujieda, M.Gertsvolf,, Y. Hanado, J. Hanssen, H. S. Margolis, G. Petit, P.-E. Pottie, C. Rieck, H. Schnatz, , A. Malimon, M. Wouters, N. Ashby, |